\documentclass[dvips]{acta}
\usepackage{supertabular,lscape,epsfig}
\usepackage{amssymb}
\usepackage{amsmath}
\usepackage{multirow}

\begin{document}

\begin{Titlepage}
\Title{On the eclipsing cataclysmic variable star HBHA~4705-03}
\Author{A.~Rutkowski$^1$, T.~Ak$^{2}$,  T.~R.~Marsh$^{3}$, Z.~Eker$^{4}$}
{
{$^1$Astronomical Observatory, Jagiellonian University, ul. Orla 171,
 30-244 Krakow, Poland\\
e-mail:a.rutkowski@camk.edu.pl \\
}
{$^2$Istanbul University, Faculty of Sciences, Department of Astronomy and Space Sciences, 34119 University, Istanbul, Turkey \\}
{$^3$University of Warwick, Department of Physics, Coventry, CV4 7AL, UK\\}
{$^4$ Akdeniz University, Faculty of Science, Department of Astronomy and Space Technologies, 07058 Akdeniz University Campus, Antalya, Turkey}
}
\Received{Month Day, Year}
\end{Titlepage}

\Abstract{We present observations and analysis of a new eclipsing binary
HBHA 4705-03. Using decomposition of the light curve
 into accretion disk and hot
 spot components, we estimated photometrically the mass ratio of the studied
 system to be $q=0.62\pm0.07$. 
Other fundamental parameters was found with modeling. This approach
gave: white dwarf mass $M_1 = (0.8\pm0.2)\ M_{\odot}$, secondary mass 
$M_2=(0.497\pm0.05)\ M_{\odot}$,
orbital radius $a=1.418\ R_{\odot}$, orbital inclination $i = (81.58\pm0.5)^{\circ}$, accretion disk radius $r_d/a = 0.366\pm0.002$, and accretion rate
$ \dot{M} = (2.5\pm2) \times 10^{18}[g/s]$, ($3\times 10^{-8}[M_{\odot}/{\rm yr}]$).
Power spectrum analysis revealed ambiguous low-period Quasi Periodic 
Oscillations
 centered at
 the frequencies $f_{1}=0.00076$~Hz, $f_2=0.00048$~Hz and $f_3=0.00036$~Hz.
The $B-V=0.04$ [mag] color corresponds to a dwarf novae during an outburst.
The examined light curves suggest that HBHA~4705-03 is a~nova-like variable star.}
{accretion, accretion discs - binaries: cataclysmic variables, stars: dwarf novae, oscillations, stars: individual: , HBHA 4705-03, 1RXS J221653.0+464804}

\section{Introduction}
Following Internet web sides of amateur astronomers we noticed a report by
Korotkii and Krachko (2006) on a new interesting
variable. They found it is an eclipsing binary. We decided to conduct
observations of HBHA 4705-03 in order to study this interesting object closer.
Light curve of this star suggests it is a member of cataclysmic 
variable stars group. 
Non-magnetic cataclysmic variable stars are short-period binaries where a white dwarf
primary is the accretor of matter from a late-type Roche lobe filling secondary star
via an accretion disk. The matter from the secondary moves into the stream and 
collides with the matter stored in the vicinity of the accretion disk.
A hot spot is
formed in the region where the stream strikes.
 
Cataclysmic variables (CVs) made a rather extensive class of objects 
characterized by a vast diversity of behaviours of their light curves 
(Warner 1995). There are {\em eruptive} and
{\em non-eruptive} members of CVs. These non-eruptive stars are called 
{\em nova-likes}. First observations of HBHA~4705-03 suggest that it belongs
to an unknown class of CVs. This previously unexplored object raised our curiosity, so we decided to investigate it.

Most recently we have found than Yakin et al. (2013) presented their work 
dedicated 
to this object. They used photometric and spectroscopic observations, while we
used only photometry. They obtained
qualitatively similar, however slightly different, result to ours.  
Since the data analysis in our study was based on a different method than Yakin et al. (2013), we decided to present our results.

\section{Observations and reduction}
Observations of HBHA~4705-03 were made on August 26 (2010). We used 
Russian-Turkish 1.5-m telescope (hereafter RTT150) located 
at the TUBITAK National
Observatory (TUG). SDSS $g'$ filter ($\lambda_0=475$nm) was used during observations. 
This telescope can work for the
photometry and the spectroscopy in two distinct modes. Namely, the coude mode 
and the cassegrain mode can be used (for detail see the official TUG webpage:
http://www.tug.tubitak.gov.tr). We made observations 
in the Cassegrain mode using ultra-fast the ANDOR iXon DU-888 CCD
camera. The iXon DU-888  camera is equipped with back-illuminated 1024x1024
pixels CCD. The instrument is mounted 
in Cassegrain focus f = 1/7.7 and gives $\sim 4 \times 4$ arcmin field of view.
This camera works in EMCCD (electron multiplaying CCD) mode, which 
reduce significantly readout noise effect at very short exposure times.
The CCD is cooled thermoelectrically to a temperature of $-60^\circ$C.
The entire CCD array can be readout up to 8 times per second. 
When reducing the readout region and binning rows, the exposure time can be
reduced to $\sim$1 ms.

We used the following procedure for the data collection (compare with Revnivtsev et al. 2012).
At first, a small sub-frame (1024x60 pixels) of the full CCD was chosen.
The variable star and a comparison star was inside this sub-frame.
After an exposure, this sub-frame is automatically binned to obtain a
"one dimensional strip" of 1024x1 pixels. In this form the exposure is saved to a hard disk.
Above procedure  allows to obtain 
exposures with the integration time of 0.0317 sec.  
The photometric measurements were made in a one-dimensional strip with a fixed
center. The aperture width was determined from the summed one-dimensional 
brightness profile on the CCD. 
During nearly 7 hours of our observational run we collected 769986
measurements of the variable star and its comparison object. 
Because of the large
number of frames collected, the data was stored in a 3-D "data cube" FITS 
file. In order to extract measurements from 
this 3-D format, we created a CCD Data
Reduction routine working under the IDL\footnote{The Interactive Data Language
(IDL) is a proprietary software system distributed by Exelis Visual
Information Solutions, Inc. (http://www.exelisvis.com)}. Intrinsic intensities 
of HBHA~4705-03 were obtained by dividing the variable star fluxes over the comparison star.
Comparison star (hereafter Comp A) is located at $\alpha=$ 22:16:47.42, $\delta=$+46:47:30.3.
To obtain the apparent magnitude $g'$, we assumed a linear change 
in spectral flux distribution between $B$ and $V$ passbands. 
Then the $g'$ of Comp A was interpolated.
$B$ and $V$ magnitudes
were taken form the Naval Observatory Merged Astrometric Dataset (NOMAD1)
(Zacharias et al. 2005) which is linked with the SIMBAD astronomical database.
This approach 
allows us to estimate the magnitude $g'= 14.8\pm0.5$ of the Comp A.

In order to better
understand the studied object we observed it on June 11,
and June 12, 2012 with the 1-m TUG telescope. Each observing run lasted
$\sim 4.5$ hours. Observations were made mostly in the $V$ filter but time 
to time we also used $B$ and $I$ filters.
This telescope has been designed for Ritchey-Chr\'{e}tien
 optical system and is equipped with the SI (Spectral Instrument) 1100 Series 
CCD camera. The whole CCD chip includes 4096x4037 pixels and is cooled by
closed cycle refrigeration unit down to -100 C
(http://www.tug.tubitak.gov.tr/t100\_si\_ccd.php).

During two nights of observations
we collected 817 frames, with the average exposure time 20 sec.
We used the standard data reduction and aperture photometry to 
analyse these data.
The object located at $\alpha=$22:16:54, $\delta=$+46:46:04 was used as
a comparison star (Comp B).

Figure 1. presents an exemplary reduced frame where the variable and the
comparison stars are marked.
\begin{figure}
\centering
\includegraphics[width=0.8\textwidth]{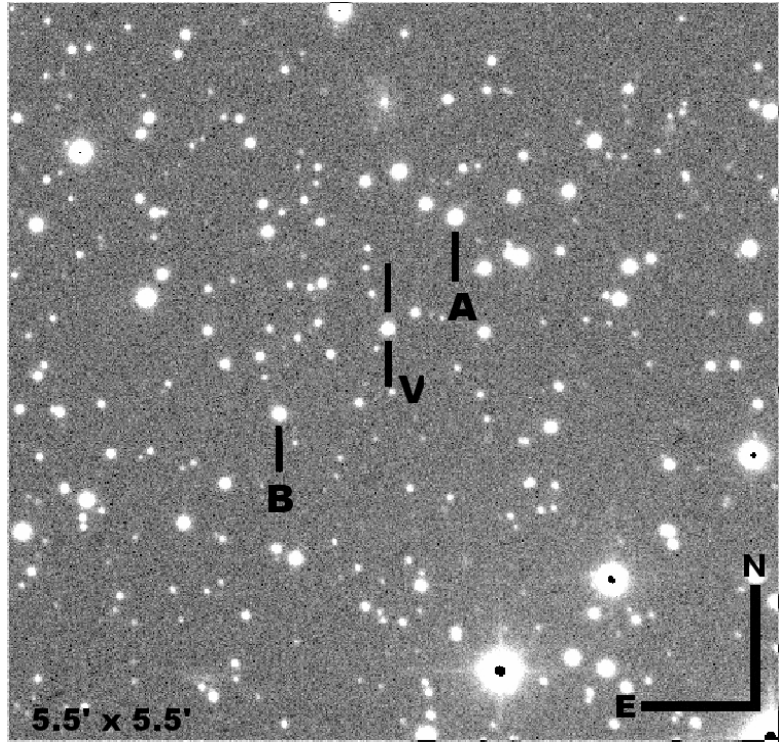}
\caption{Finding chart of HBHA~4705-03. The variable is marked by the V. 
The comparison star (A) was used for the RTT150 telescope data, while the comparison star (B) 
was used for the data gathered in 2012 with 1-m telescope. The chart gives 
roughly 5.5'x5.5' field of view.}
\end{figure} 

\section{Light curve}
\subsection{Data from  August 26, 2010 obtained with RTT150}
We collected 769986 frames in the 'g'' filter. 
The raw photometric data do not reveal brightness modulation because of a high
noise level. Due to the large number of points we binned them with 20 
second bands to increase the S/N ratio.
The obtained light curve
can be found in Figure 2. Average errors for these observational points are 
close to 0.04 mag.
\begin{figure}
\includegraphics[width=0.5\textwidth,angle=-90]{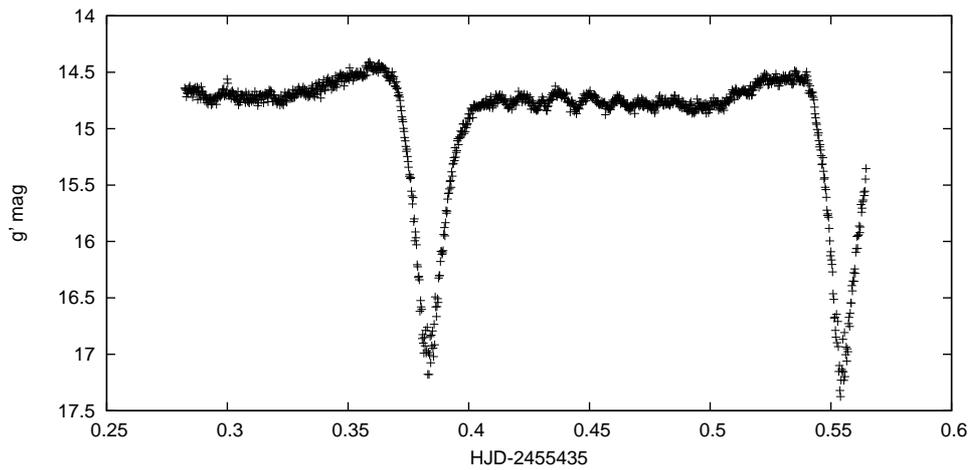}
\caption{Light curve of HBHA~4705-03 in the $g'$ filter obtained with the RTT150 on August 26, 2010}
\end{figure}
\\

\noindent We detected two prominent minima with the amplitude
 $\sim 2.5$mag.
Before each of the eclipses a prominent hump can be observed. This is a well known profile
for eclipsing cataclysmic variable stars. The mentioned humps are obviously 
the manifestation of the hot spot located on the disk - stream intersection.
The orbital
revolution causes reorientation of the hot spot and in consequence variations 
the  in the observed light curve.
In addition, we can notice a prominent, intriguing tooth-shape modulation
in the middle between consecutive minima. We do not have a simple interpretation for
this behaviour and we will return to this phenomenon in Section 4.
\subsection{Data from June 11 and 12, 2012 obtained with 1-m telescope}
Figure 3 presents the light curve obtained on June 11 and 12 in 2012.
\begin{figure}
\includegraphics[width=0.7\textwidth,angle=-90]{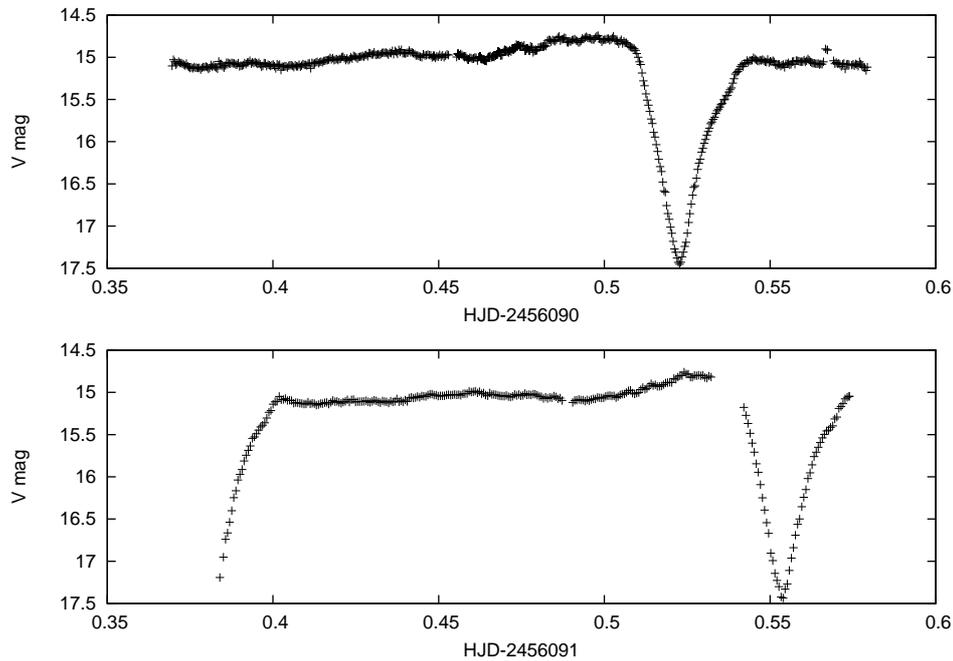}
\caption{Light curve of HBHA~4705-03 in the $V$ band obtained in June 11 and 12 in 2012 with the 1-m telescope.}
\end{figure}
During this run we used only the $V$ filter to obtain the light curve, while $B$ and $I$ 
measurements were made only one time in order to check the color 
characteristic. 
The $B$ and $V$ magnitudes of the comparison star B were taken from the NOMAD1
catalog, while the $I$ magnitude was taken from the USNO-B1 catalog.
After calibration we estimated magnitudes of HBHA 4705-03 to be 
15.25 in $B$,15.21 in $V$ and 14.03 in $I$ (the measurements were obtained close to date
HJD=2456090.5661). 
The obtained light curve presents a very good
quality and in general the errors are around 0.005 mag.
Again, there are visible obvious eclipses caused by 
transits of the secondary which covers the accretion disk and the hot spot.

However, the tooth-shape modulation occurring between eclipses in the RTT150
data from 2010 are not visible in the data obtained with 1-m telescope in 2012.

  Either this modulation is not
present in the system during the 2012 run of observations or they are simply not visible in the $V$ band.
A temporal appearance and disappearance of these tooth-shape variations would
not be unusual. 
Many other variable stars belonging to the sub-class of novae or nova-like
 objects change their curves in the similar way to HBHA.
Our experience 
tells us that the light curve variations can affect both its 
shape and the brightness.
\subsection{Orbital period of the system}
One of the most important physical parameter which defines many others 
in cataclysmic variable star is
the orbital period. In the case of HBHA~4705-03, we have only 4 determined eclipses over two years. This is simply not enough
for the standard Fourier analysis. Instead, we determined times of the minima by fitting a polynomial function
to each of the eclipses. Then we found periods between detected minima in each year of observations. After calculation of the average of the results from
2010 and 2012, we found that the period equals to
0.171867(59) days. We interpret this value as the orbital period of the 
system.
\section{Frequency analysis of the RTT150 data}
High time resolution of the RTT150 data gives us opportunity to conduct the
frequency analysis. As a tool we used the ZUZA code written by 
Schwarzenberg-Czerny (1992). The orbital humps after maxima and eclipses were
excluded from the analysis. We examined two ranges of the light curve, i.e. 
between $0.27-0.33~\rm{d}$ and $0.41-0.51~\rm{d}$ (see Fig 2).
The obtained power spectrum is shown in Figure 4.
\begin{figure}
\includegraphics[angle=-90,width=0.9\textwidth]{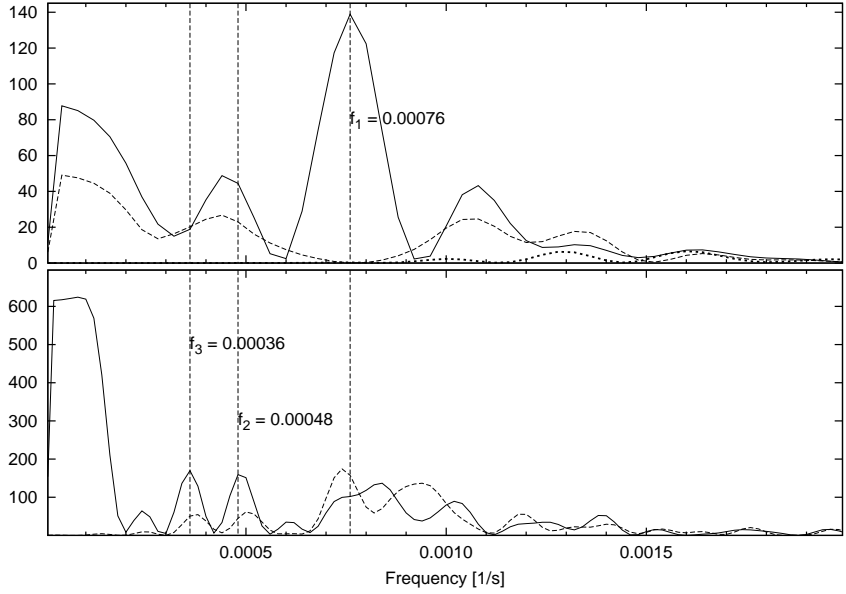}
\caption{Power spectrum for the RTT150 data. Upper panel:
the periodogram for the data before the first minimum, i.e.
between HJD = 2455435.27 and 2455435.33. Bottom panel: periodogram for the data from
observation period between HJD = 2455435.41 and 2455435.51. See text for 
detailed description.
}
\end{figure}
We used the {\em perort} routine of ZUZA to obtain the AOV (Analysis of Variance - Schwarzenberg-Czerny 1989) 
periodogram with two harmonics.
The periodogram for the
data collected before the first maximum (the upper panel on Fig.4) shows
a prominent peak centered at the frequency $f_1=0.00076(7)$~Hz (solid curve).
Doubled prewhitening procedure allowed us to remove low-frequency variations from the light curve. Resulting periodograms shown by dashed and dotted lines in the Fig. 4
present no other significant frequency.
The power spectrum of the second range of data is presented on bottom panel of Fig 4.
Here the situation is different. Much more complex structure of the peaks 
can be noticed. Neglecting low-frequency signal in the spectrum 
which is resulting from the imperfectly removed trend, we can measure two peaks at the
frequencies $f_2=0.00048(3)$~Hz and  $f_3=0.00036(3)$~Hz. After prewhitening
and removing low-level frequency peak centered at the frequency
 $8\times10^{-5}$~Hz
with its three harmonics we got the spectrum presented by the dashed line. Thus, we 
revealed an additional frequency close to $f_1$ which is also present in the data before the first 
minimum. It suggests that Quasi-Periodic Oscillation (QPO) around $f_1$ is
persistent
for the whole observational period -- before and after the minimum. 
An ambiguous characteristic of detected variability (in particular for long periods)
do not allow them to be classified as Dwarf Nova Oscillations (DNOs) neither
to common QPOs. However,
this type of variability -- in analogy to the known characteristic of DNOs and
QPOs in dwarf novae (Warner \& Woudt 2005) -- suggests a high mass transfer rate 
during the time of observations.
\section{Analysis of the eclipses}
\subsection{Decomposition into spot light curve and disk light curve}
Detected eclipses in the light curve gave us a chance to find
fundamental physical parameters of HBHA~4705-03. The most fundamental parameter is the orbital period of the system which we estimated in Section 3.3. 
To derive other parameters, a more developed method must be used.
After the intensive study of literature dedicated
to the problem of analysis of eclipse light curves of CVs (see the review of Horn 1993) we decided to follow the approach by Smak (1994a,b).
Introducing minor modifications into the orginal method we have decomposed
 the observed light curve on the hot spot light
 curve and the disk light curve. This analysis gives the best result for 
the shape of eclipse light curve which is described by the standard model of
a cataclysmic binary with stationary accretion. For short,
this model predicts approximately constant brightness of the system during 
roughly one-half of the cycle and orbital hump, which occurs in the phase
interval from about $\phi=-0.4$ to about $\phi=0.1$. Particularly important 
from the point of view of the assumed methodology is the fact that the declining part 
of the orbital hump is still clearly visible after the eclipse.

Analysis of the light curve from 2010 did not give
satisfactory results.  Evidently the shape of the eclipse which is far 
from the "standard" case made it impossible to obtain the reliable result.
Thus, we decided to obtain an additional observations of the studied object. We used the
data gathered from the 2012 campaign to produce phased eclipse with minimized
flickering.  Results of this approach are presented in the Fig. 5.
One can notice a clear shape of the hot spot light curve. Four phases
of contact ($\phi_1, \phi_2, \phi_3,\phi_4$) can be determined with
relatively good accuracy which is estimated
as $\pm 0.003$ in phase units. Those phases are marked by vertical lines in the Fig. 5.
Although, the presence of the flickering is still visible
in the light curve, the shape of the uneclipsed part of hot spot light curve
agrees quite well with the theoretical formula presented in Paczy\'nski \& Schwarzenberg-Czerny(1980, Eq.4)
\begin{equation} 
l(\phi)=l_{s,max}[1-u+u\cos(\phi-\phi_{max})]\cos(\phi-\phi_{max}).
\end{equation}
Used symbols denote: $l_{s,max}$ - the hump amplitude, u is the limb darkening
coefficient (for which we adopt $u = 0.6$) and $\phi_{max}$ is the phase of
hump maximum.
Table 1. presents the fitting parameters of this function:
\MakeTable{c|c|c|c|c|c}{12.5cm}{Hump parameters and phases
of contacts for the hot spot fitting in HBHA~4705-03}{\hline
$l_{s,max}$ & $\phi_{max}$ & $\phi_{1}$ & $\phi_{2}$ & $\phi_{3}$ & $\phi_{4}$\\
\hline
0.382 & -0.052 & -0.078 & -0.035 & 0.093 & 0.107 \\
\hline
}
\begin{center}
\begin{figure}
\includegraphics[width=1.0\textwidth]{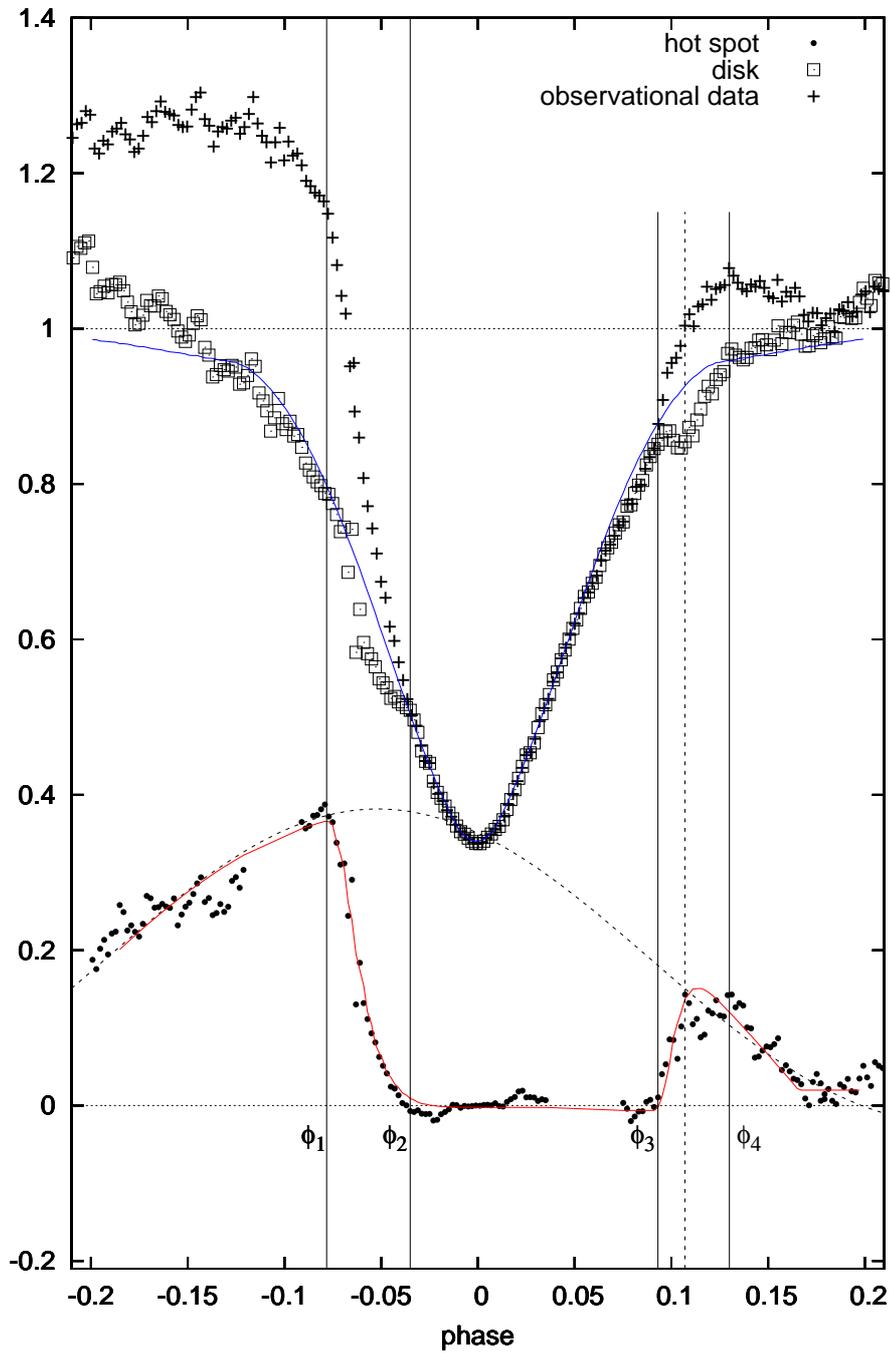}
\caption{Decomposition of the HBHA~4705-03 light curve into spot and disk
components. The observed light curve is shown by crosses. Dots indicate 
reconstructed spot light curve.
The theoretical spot light curve outside the eclipse are 
presented by dashed line. The solid curves are the
{\em model} light curves corresponding to the disk light solution (upper) and the spot light curve solution (bottom).  Vertical lines indicate 
four moments of eclipse contacts.}
\end{figure}
\end{center}
Reconstructed disk light curve is also well defined. However, its shape is not
clearly symmetric. Most likely two effects play a role here: flickering and 
the complex structure of the disc, especially non uniform distribution of 
the surface brightness. We should also take into account that it may 
not be the case of stationary accretion. Even so, we do not have much
more other ways to estimate the mass ratio of the system. Despite the danger of
contradiction between the possible non-uniformity of the disk and the 
assumption of the stationary accretion
we hope at least for crude estimation of the $M_1$ and $q$.
\subsection{Model analysis and assumptions}
One of the most serious issues regarding light curve model analysis 
is the problem 
of appropriate assumptions. One have to find the way to collect as many
starting parameters as possible. In Sec. 3.3, we derived the first one -- orbital period. 
The second
parameter which affects the other parameters is the stellar components mass
 ratio. Here we
used the convention $q=M_2/M_1$, where $M_2$ is the mass of the donor and $M_1$
is the mass of the white dwarf.
Following Smak's approach (Smak 1994) we did not assume value of the mass
ratio $q$ but instead, we check a wide range of parameters in order to
estimate it.
Knowing masses and the orbital period we can derive 
the orbital semi-major axis $a$ {\em via} the Kepler Law.

Cataclysmic stars obey the well know relation between the orbital period
and the mass of the secondary, $M_2=f(P,M_1)$  Patterson (1984). This allowed
us to estimate the mass of the secondary to be $M_2/M_\odot=0.4970$.
The temperature of the secondary was derived with the following formula taken from Popper (1980):
\begin{equation}
\log T = 3.760 + 0.633 \log M_2/M_{\odot}
\end{equation}

We assumed that the relative luminosity of the white dwarf (WD) is small and 
WD itself is obscured by internal regions of the accretion disk. Based also on
the other works (e.g. Sion 1991) we adopted here the standard temperature
of the white dwarf which is suitable for such a system, $T_1=30000$ K. At this point,
we need to add that the final fit barely depends on the assumed 
temperature. 
The radius of the primary can be determined from the theoretical mass radius
relation (e.g. Nauenberg 1972, Provencal et al. 1998) for white dwarfs.
The surface brightness distribution of the component in a given
passband is a function of temperature and limb darkening $u$. Here we adopt
the standard value of the linear limb darkening coefficient, $u = 0.6$, for WD and disk components. For the secondary we assumed that this coefficient is equal to 0.3. 
We also assume the standard temperature profile for a flat disk
$T_{d}\sim R^{-3/4}$.

Therefore, the starting point for the modeling was the following set
of parameters:
$M_1$, $M_2$, $R_1$, $a$, $u_1$, $u_2$, $u_{disk}$, $T_1$, $T_2$. 
We also assumed that the secondary fills its Roche lobe and the hot spot is located
on the edge of the accretion disk.
As a useful tool for modeling we used the LCURVE program written by Tom Marsh. 
\begin{figure}
\centering
\includegraphics[angle=-90,width=0.9\textwidth]{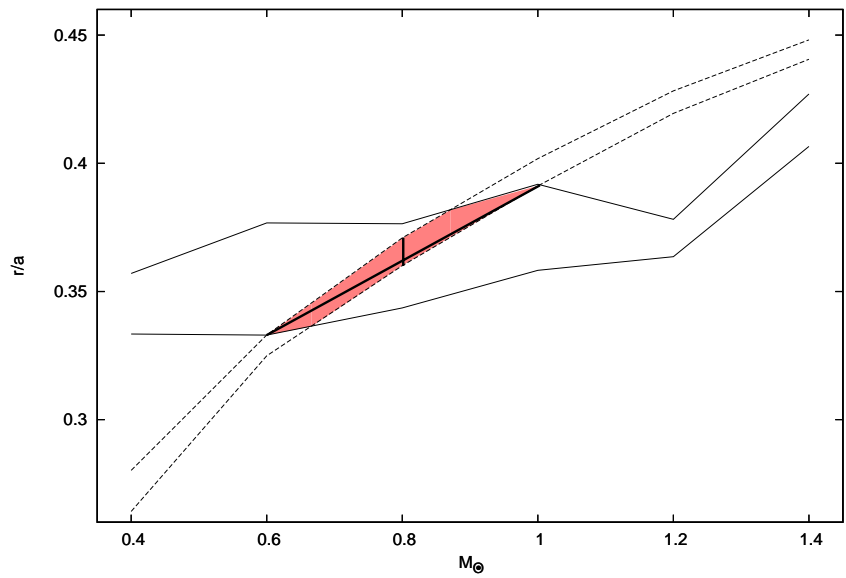}
\caption{The mean $r_d/a = f(M_1)$ relation from disk light solutions is shown by
 solid lines. The upper and the bottom lines show $r_d/a$ determination errors. 
Dotted lines represent $r_s/a = f(M_1)$ relation for spot light solution
($r_d$ represent the outer radius of the disc and $r_s$ is the distance of the hot
spot to the white dwarf).
Oblique cross shows the final solution for $M_1$ and $r_d/a$. Its possible
uncertainty is presented by the shady region.}
\end{figure}
Decomposition of the light curve allowed us to analyze the disk and the
hot spot components separately. 
Starting from the modeling of the disk light curve we have tested different
 masses of the primary (from 
$0.4M_{\odot}$ to $1.4M_{\odot}$). Here we modeled
 only the inclination of the system, the radius, and the temperature of the accretion disk.
Based on the knowledge on vertical structure of the disk, we assumed the most
probable value of the disk thickness. The average mass transfer rate for a nova-like variable
was found to be $\sim9.3\times10^{-9} \rm{M_{\odot} yr^{-1}}$ (e.g. Puebla et al.
2007, Ballouz et al. 2009). Using the analysis made by Smak (1992) we
 adopted $z/R \approx 0.063$ in our models.
When testing this assumption, we noted
that changing $z/R$ in a rather wide limit (from 0.63 to 0.1) 
change the system inclination noticeably. On the other hand the disk radius
is changed barely.

Similar modeling was conducted for the spot light curve. We assumed that the
radius of the accretion disk is the same as the distance of the white dwarf
to the hot spot, i.e. the hot spot is located on the accretion disk edge.
 Initial parameters for this step of analysis
(like the temperature of the accretion disk and the system inclination)
was taken from the disk light curve solution.

Both model analyses described above give a family of solutions for
 different assumed $M_1$.
Figure 6 presents the 
relations of the accretion disk radius to semi-major axis ratio versus the
the primary mass obtained both from the disk light curve solution ($r_d/a=f(M_1)$) and the spot light 
curve solution ($r_s/a = f(M_1)$) presented by the solid and dotted lines,
respectively. Placing both families of the solutions on one graph we obtained 
a shady region presenting possible values of $M_1$.
The black cross indicates the center of the found zone and the most likely
value of $M_1$ and $r_d$. 
Moreover, the obtained synthetic disk and spot light curves are presented in 
Figure 5. Turning our attention to disk light curve 
we can notice that the fit for is very good in the center, however, less perfect in the ''wings''. Systematically fainter {\em model} disk light curve suggests 
the higher temperature of the outer regions of the accretion disk. Not strictly 
smooth observed disk light curve suggests a non uniform structure of the disk
which is 
very difficult to model. The resulting model of the spot light curve seems  
very good, though. This model differ very little from the theoretical relation given by Paczy\'nski \& Schwarzenberg-Czerny (1980).
\section{Results}
Disk/spot light curve solutions provide the following parameters for the system
$M_1/M_{\odot}=0.8\pm0.2$, $M_2/M_{\odot}=0.497\pm0.05$, $a=1.41821 \rm{R_\odot}$, 
$r_d/a=0.36\pm0.02$, $i^{\circ}=81\pm0.1$.
Those parameters were taken as initial parameters used later for the overall
observational light curve modeling. 
LCURVE code allowed us then to model HBHA~4705-03 including four components:
the white dwarf,
the secondary star, the accretion disc and the hot spot created at the place
where the stream hits the edge of the disk.

The mass ratio was fixed to be $0.62$. The rest of the parameters (except those assumed
in section 4.2) were obtained using model analysis. 
The best fit of the light curve solution in this approach is presented 
in Fig. 7. Generally,
the synthesized light curve agree quite well with the observed light curve
of the eclipse.
Table 1 presents parameters of the best fit solution and Fig. 7 shows its
 graphical representation. Simplex minimization method was used
 at first to find the global minimum. Next, the Levenberg-Marquardt method
(Levenberg 1944, Marquardt 1963, see also Bates 1988) was used 
to obtain the final solutions by the {\em levmarq} routine. 
Once the optimal curve-fit
parameters were determined, their errors were deduced from the
parameters the covariance matrix.
\begin{center}
\MakeTable{ c c c }{12cm}{Resulting parameters of HBHA~4705-03}{
  \hline                        
  $M_{1}/M_{\rm \odot}$ & = & $0.8\pm0.2$ \\
  $M_{2}/M_{\rm \odot}$ & = & $0.497\pm0.05$ \\
  a [$R_{\odot}$]       & = & 1.41821 (calculated)\\
  $i[^{\circ}]$         & = & $81.58\pm0.5$ \\
  $t_1[K]$      	& = & 30000 (assumed) \\
  $t_2[K]$	        & = & 3696 (calculated) \\
  $r_d/a$		& = & $0.366\pm0.002$ \\
  $\dot{M} [g/s]$       & = & ($2.5\pm2) \times 10^{18}$, ($3\times 10^{-8} M_{\odot}/{\rm yr}$) \\ 
  \hline  
  }
\end{center}

\begin{figure}
\centering
\includegraphics[angle=-90,width=1.0\textwidth]{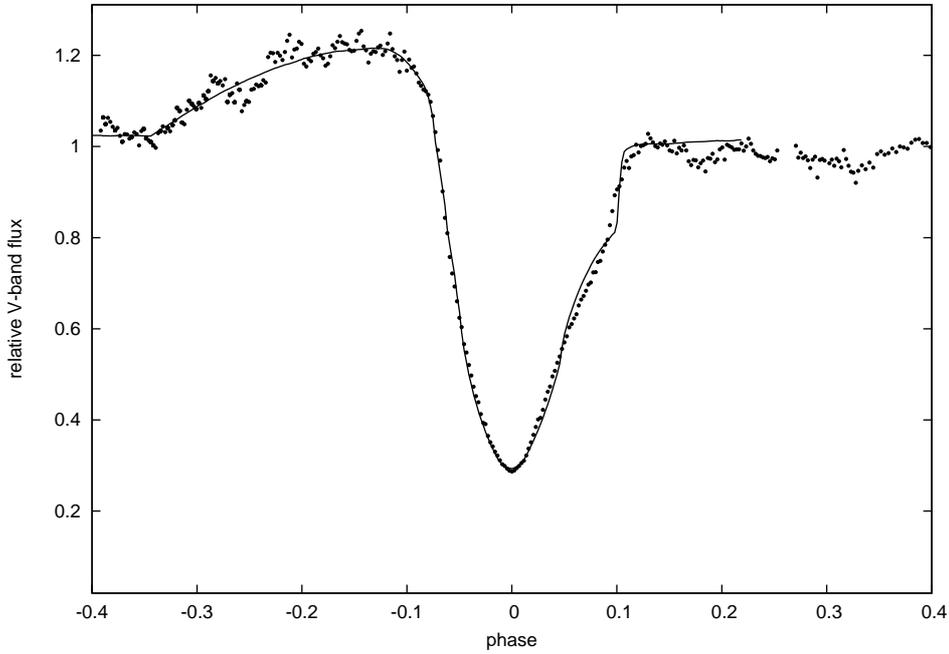}
\caption{Comparison between the observational data of HBHA~4705-03 and the best fit found with the LCURVE code (solid line)}
\end{figure}
\subsection{Comparison with Yakin et al.}
Comparison of our results with those presented by Yakin  et al. (2013)
suggest that one should treat both results with caution.
The main problem in the modeling of the light curves is the correct estimation
of the mass ratio of the components and the inclination of their orbit plane.
 Theoretical relations allow
to estimate $M_2$ with statistical good accuracy. So, the main parameter
to be determined is the white dwarf mass.
We found $M_1=0.8\pm0.2 M_{\odot}$. The error of this measurement gives
a wide possibility of the modification for the other parameters.
For example, if we adopted $M_{1}=0.6 M_{\odot}$, then this value would agree
with $M_{1_{Y}}=0.54\pm0.1 M_{\odot}$ found by Yakin et al. The rest of the parameters
would change accordingly to the mass ratio.
One should also remember about the six-year time difference between the 
observations of Yakin et al. and ours. In particular, for that reason, the disk
and the hot spot parameters can be significantly different and 
can bias the rest of the parameters obtained from the model analysis.
\section{Conclusions}
Part of the below conclusions was already introduced in the abstract of this 
article. 
The observations of the new discovered variable star HBHA~4705-03 in the years
2010 and 2012 were presented. We were able to decompose the observed light curve into
the hot spot and the accretion disk components. The eclipse of the hot spot is
clearly visible. Model analysis of the both of light 
curve components separately allowed us to find intrinsic parameters for the disk
and the spot. After combining these two solutions, we were able to find the mass ratio of the
system and the accretion disk radius. At this point we had enough data to find the 
 remaining global parameters. 
We detected uncommon variability with the period in the range from $\sim$20 
to $\sim$50 minutes.
Moreover, the LCURVE code allowed us to find 
the overall synthetic light curve of the system, including all major components i.e. the white dwarf, the red dwarf, the accretion disk and the hot spot. 
\subsection{Classification of HBHA~4705-03}
The shape of the light curve and the found parameters like
the orbital period,
the accretion rate, the component masses and the $B-V$ color, help us to
made a classification of this object. Moreover, although there are no long-term light curves, eclipse light curve and spectra in this study combine with
Yakin et al. (2013) may be very helpful to find the type of HBHA~4505-03.

As the orbital period of HBHA~4705-03 is very close to 4 hr, we focus on the 
CVs with period between 3 and 4 hr which are right at the upper edge of 
the period gap. In this period range, there are a lot of nova-like stars
characterized by an approximately steady, high rate of mass transfer. In
addition, nova-like stars with
orbital periods between 2.8 and 4 hr are classified as SW Sex stars. These stars were first defined by Thorstensen et al. (1991). Observational features of
SW Sex stars are summarized by Hoard et al (2003). They are high mass transfer
rate nova-like stars. In the optical light curve of SW Sex stars, the white
dwarf and the accretion disc are deeply eclipsed by the secondary star.
This shows
that their orbital inclination angles is higher than 80$^{\circ}$. However,
there are non-eclipsing SW Sex stars (Schmidtobreick et al. 2012), as well.
They display high excitation
spectral lines, including He II $\lambda$4686 emission, which strength is
often comparable to the strength o H lines. They show single-peaked emission
lines in their spectra rather than double-peaked ones expected from
a near-edge-on accretion disc. The Balmer and He I emissions are only shallowly eclipsed compared with the continuum emission. The zero crossings of their
emission-line radial velocities present phase offsets relative to their eclipse
ephemerides.

At first glance HBHA~4705-03 can be considered as a nova-like
star since its mass transfer rate is very high compared to dwarf nova. 
In addition, this system exhibits many features similar
to SW Sex stars. White dwarf and accretion disc of the system is deeply
eclipsed by the secondary star. This is evident as its inclination angle
is found about 81$^\circ$ in our study. Optical spectra of HBHA~4705-03 are presented by Yakin et al 2013. In its optical spectra, He II$\lambda$ emission
line is prominent and its strength is comparable to H lines.  Emission lines in
its spectra are single-peak and the zero crossing of the radial
velocity of He I$\lambda$4921
emission line shows a phase offset comparing to its eclipse
ephemerides. These features demonstrate that HBHA~4705-03 can be a member of SW Sex class of the nova-like-type CVs.

\Acknow{The project was supported by Polish National Science
Center grant number DEC-2012/04/S/ST9/00021 awarded to AR.
We would like express our gratitude to Excellence Cluster Universe,
Technical University, Munich for purchase of the iXon CCD camera.
This work was also supported by the TUBITAK Programs 209T055 and 09ARTT150-427.
The authors would like to thank I. Khamitov and A. Tkachenko for their
help during the observations. We thank Stanis\l{}aw Zo\l{}a, Wac\l{}aw Waniak 
and Magda Otulakowska-Hypka for a careful reading of the manuscript and their
useful suggestions.}

\end{document}